\documentstyle[twoside,fleqn,espcrc2]{article}

% ...
\newcommand{\VEV}[1]{\left\langle #1\right\rangle}
\renewcommand{\Re}{{\rm Re}}
\renewcommand{\Im}{{\rm Im}}
\newcommand{\Tr}{{\rm Tr}}

\gdef\journal#1, #2, #3, 1#4#5#6{		
    {\sl #1~}{\bf #2} (1#4#5#6) #3}		
\def\prd{\journal Phys. Rev. D, }
\def\prl{\journal Phys. Rev. Lett., }
\def\np{\journal Nucl. Phys., }
\def\pl{\journal Phys. Lett., }
\begin{document}
\title{Baryon Density Correlations in the Quark
Plasma\footnote{Presented by C. DeTar}}

\author{MIMD Lattice Calculation (MILC) Collaboration:
C.~Bernard,\address{
Department of Physics, Washington University, St.~Louis, MO 63130, USA}
T.~A.~DeGrand,\address{
Physics Department, University of Colorado, Boulder, CO 80309, USA}
C.~DeTar,\address{
Physics Department, University of Utah, Salt Lake City, UT 84112, USA}
S.~Gottlieb,\address{
Department of Physics, Indiana University, Bloomington, IN 47405, USA}
A.~Krasnitz,$^{d}$
R.~L.~Sugar\address{
Department of Physics,
University of California, Santa Barbara, CA 93106, USA}
and D.~Toussaint\address{
Department of Physics, University of Arizona,
Tucson, AZ 85721, USA}
}

\begin{abstract}
      As part of an ongoing effort to characterize the high
temperature phase of QCD, we measure the quark baryon density in the
vicinity of a fixed test quark and compare it with similar
measurements at low temperature and at the crossover temperature.
Such an observable has also been studied by the Vienna group.  We
find an extremely weak correlation at high temperature, suggesting
that small color singlet clusters are unimportant in the thermal
ensemble.  We also find that at $T = 0.75 T_c$ the induced quark
number shows a surprisingly large component attributable to baryonic
screening.  A simulation of a simple flux tube model produces results
that suggest a plausible scenario: As the crossover temperature is
approached from below, baryonic states proliferate.  Above the
crossover temperature the mean size of color singlet clusters grows
explosively, resulting in an effective electrostatic deconfinement.
\end{abstract}

\maketitle

\section{MOTIVATION}
Numerical simulations of the quark plasma have suggested seemingly
contradictory models.  While bulk thermodynamic quantities, such as
the energy density\cite{ref:endens} and baryon
susceptibility\cite{ref:barsus} yield values consistent with a nearly
free gas of quarks and gluons, measurements of screening propagators,
particularly, measurements of the wave functions of exchanged objects,
reported in Lattice '91, are consistent with the confinement of color
singlets\cite{ref:screen_wavef}. Indeed, simulations and analytic work
in the pure glue sector have demonstrated that space-like Wilson
loops obey an area law in the high temperature phase, a signature of
confinement\cite{ref:Borgs_PM} .

One resolution of this seeming paradox describes the quark plasma as
an ensemble of color singlet clusters of various sizes.  Bulk
thermodynamic quantities, such as the energy density would receive
contributions from all clusters, whereas long-range screening would be
controlled by the lightest clusters.  How large is the typical color
singlet cluster?  What is the typical spatial extent and quark and
antiquark content?  To answer this question, it is necessary to seek
observables that have not hitherto been studied in this context.  Thus,
we measured the distribution of induced quark charge (baryon number)
in the vicinity of a fixed test quark, at low and high temperature,
and at the crossover temperature.  This observable has also been
studied by the Vienna group\cite{ref:FFM}. At low temperature we
expect that, as a result of confinement, a dynamical antiquark or, less
often, a pair of quarks, screens the test charge at short distance.
Thus, the induced dynamical quark number density should be large and
negative close to the test charge.  If screening is entirely due to a
single antiquark, we should observe that the total induced quark
number $Q$ is $-1$.  By contrast, if color singlet clusters are large
both in size or in the number of quarks and antiquarks, we would
expect only a weak correlation and a small value of $Q$.

\section{QUARK NUMBER DENSITY}
The construction of the local quark number density starts with the
introduction of a baryon chemical potential in the standard
way\cite{ref:chem_pot}, but with a spatial dependence\cite{ref:paper}.
Such a definition assures that the total baryon charge so defined is
exactly conserved on the lattice.  Differentiating the thermodynamic
potential with respect to the local chemical potential yields the
local quark number density.  In the staggered fermion formalism, the
quark number density (including all flavors) in the presence of a
fixed quark in the ensemble is given by the correlation:
\begin{equation}
 \rho_q({\bf r}) = -(N_f/2)\frac{
	\VEV{\Im P_{stat}(0)\Im P_{dyn}({\bf r})}_U}{\VEV{\Re P_{stat}(0)}_U},
\end{equation}
where the Polyakov loop is given by the color trace (including the
staggered Dirac phase factors $\eta_{({\bf r},t);t}$)
\begin{equation}
  P_{stat}({\bf r}) = \Tr_c\left[\prod_{t=0}^{N_t}
	\eta_{({\bf r},t);t}U_{({\bf r},t);t}\right],
\end{equation}
and the dynamical quark charge density is given in terms of the
fermion propagator $M^{-1}_{r,r^\prime}$
\begin{equation}
	P_{dyn}({\bf r}) = \eta_{({\bf r},0);t}
	\Tr_c\left[M^{-1}_{({\bf r},1),({\bf r},0)}U_{({\bf r},0);t}\right].
\end{equation}
An alternative dynamical density operator averages over all time
slices.

\section{RESULTS OF SIMULATIONS}

Simulations were carried out at fixed $\beta = 5.445$ and quark mass $m
= 0.025$ for two flavors of staggered fermions on lattices of size
$16^3 \times N_t $, where $N_t = 8,6,4 $.  This choice of lattice
parameters corresponds to the crossover temperature at $N_t = 6
$\cite{ref:Nt6}.  Thus, the simulations are done at three temperatures
$T = 0.75 T_c $, $T = T_c$, and $T = 1.5 T_c$, respectively, at the
same lattice scale.  Spectroscopic simulations at the same
temperature\cite{ref:Gottlieb} allow us to set the scale, viz.\ $T_c
= 145 $ MeV and $a = 0.227 $ fm.  The quark number density
was obtained using a random source technique, and the correlation
convolution was constructed with the aid of a Fourier transform.

 \begin{figure}[htb]
 \makebox[55mm]{\rule[-30mm]{0mm}{60mm}}
 \caption{Quark number density induced by a fixed quark at the origin
at three temperatures.  Curves are fits to a single screening mass.
The total induced charge $Q$ is obtained by integrating the fit.  }
 \end{figure}

Figure~1 summarizes our preliminary results for the induced quark
number density at these three temperatures.  Particularly striking is
the dramatic decrease at high temperature.  The total induced quark
number $Q$ is also indicated in the figure, normalized to one for a
single quark.  At the high temperature point the total induced charge
is nearly two orders of magnitude smaller that the charge at low
temperature.  Thus, we see no evidence for small color singlet clusters
in the high temperature plasma.  At low temperature we expect that the
test charge is attached to a color singlet cluster.  A single
antiquark would contribute $-1$ to the total induced charge, and a
pair of quarks, $+2$.  Evidently, already at a temperature of $0.75 T_c$,
there is a significant baryonic screening component.

\section{FLUX TUBE MODEL}

Some years ago Patel\cite{ref:fluxtube} proposed a flux tube version
of the three-state three-dimensional Potts model to explain the
mechanism of the deconfining phase transition in QCD\@.  In this
model, each site $\bf r$ of a cubic lattice holds either a quark,
antiquark, or none at all, and each link $\ell_{\bf r,\mu}$, a triplet
or antitriplet flux, or none at all.  Flux is conserved modulo 3.  The
hamiltonian is given in terms of the quark mass $m$ and the string
link energy $\sigma$ by
\begin{displaymath}
   H = \sum_{\bf r,\mu} \sigma |\ell_{\bf r,\mu}| + \sum_{\bf r}m |n_{\bf r}|.
\end{displaymath}
Patel proposed using this model as a paradigm for the QCD phase
transition.  It has some intriguing features.  At low temperature,
only small color singlet clusters may occur in the Gibbs ensemble.  As
the temperature is increased, clusters of increasing size populate the
ensemble.  Eventually clusters connect to fill the entire spatial
volume.  For heavy quark masses this phenomenon corresponds to a
first-order deconfinement phase transition.  For light quarks, as seen
in Fig.~2, simulations show that cluster growth is explosive,
 \begin{figure}[htb]
 \makebox[55mm]{\rule[-30mm]{0mm}{60mm}}
 \caption{Cluster size vs inverse temperature for the flux tube model}
 \end{figure}
eventually filling the entire spatial volume.  Here a cluster (color
singlet) is defined as a set of sites connected by flux tubes.  Shown
is the mean fraction of the total volume occupied by the cluster
connected to the origin for a $10^3$ lattice with $m = \sigma = 1$.
The vertical bar indicates the approximate crossover $T_c$.
(Curiously, despite the pervasive growth of the mean cluster size,
there is no evident accompanying ``percolation'' phase transition at
this quark mass.)  Thus, the addition of a single test quark at high
temperature in such an ensemble produces an insignificant
perturbation, thereby accounting for the extremely weak correlation
seen at high temperature in the QCD simulation.  We have an effective
electrostatic deconfinement without a phase transition.  Thus, our
results for the induced charge in QCD may find an explanation
within this picture.

The total induced charge $Q$ in the flux tube model can also be
observed.  Here, too the value is surprisingly less than $-1$ in
magnitude at $T = 0.78 T_c$.  From Fig.~3 we find that $Q = -0.67(9)$
at the same parameter set and lattice size as Fig.~2.  A visual
examination of the lattices as in Fig.~4 shows why.
 \begin{figure}[htb]
 \makebox[55mm]{\rule[-30mm]{0mm}{60mm}}
 \caption{Induced charge vs inverse temperature for the flux tube model.}
 \end{figure}
 \begin{figure}[htb]
 \makebox[55mm]{\rule[-30mm]{0mm}{60mm}}
 \caption{Typical flux tube lattice at $T \approx 0.78 T_c$}
 \end{figure}
In this periodic
lattice the lines denote flux, the crosses, quarks, and the octagons,
antiquarks.  This typical lattice contains one baryon, one antibaryon,
and nine mesons; by contrast a naive application of Boltzmann
statistics to only the lowest lying meson and baryon in this model
would have predicted fewer than one baryon per hundred mesons.  The
baryons in Fig.~4 are not the lowest lying states.  We note that the
density of baryonic states grows with mass more rapidly than that of
mesonic states in this model.  Thus, baryonic clusters proliferate as
the temperature rises through $T_c$, permitting more frequent baryonic
screening of a test charge.

To be sure the flux tube model omits many features of QCD\@.  It lacks
dynamics, describing only electrostatics.  Completely omitted are the
important magnetic interactions that give rise to confinement
in spacelike propagation.  It would be useful to find an elaboration
of the model more closely relevant to QCD\@.  Nonetheless, it is highly
suggestive both for further exploration of QCD and for the
phenomenology of the quark plasma.

\medskip
Code development and testing were carried out on the nCUBE and Intel
iPSC/860 at the San Diego Supercomputer Center.  Computations in QCD
were carried out on an Intel iPSC/860 at the NASA-Ames Research Center
and on a Thinking Machines Corporation CM-5 at the Pittsburgh
Supercomputer Center.  We are extremely grateful for the support of
these three centers.  Flux tube model calculations were carried out on
IBM RS/6000's at the University of Utah.  This work was supported by
the U.S.\ Department of Energy and by the National Science Foundation.

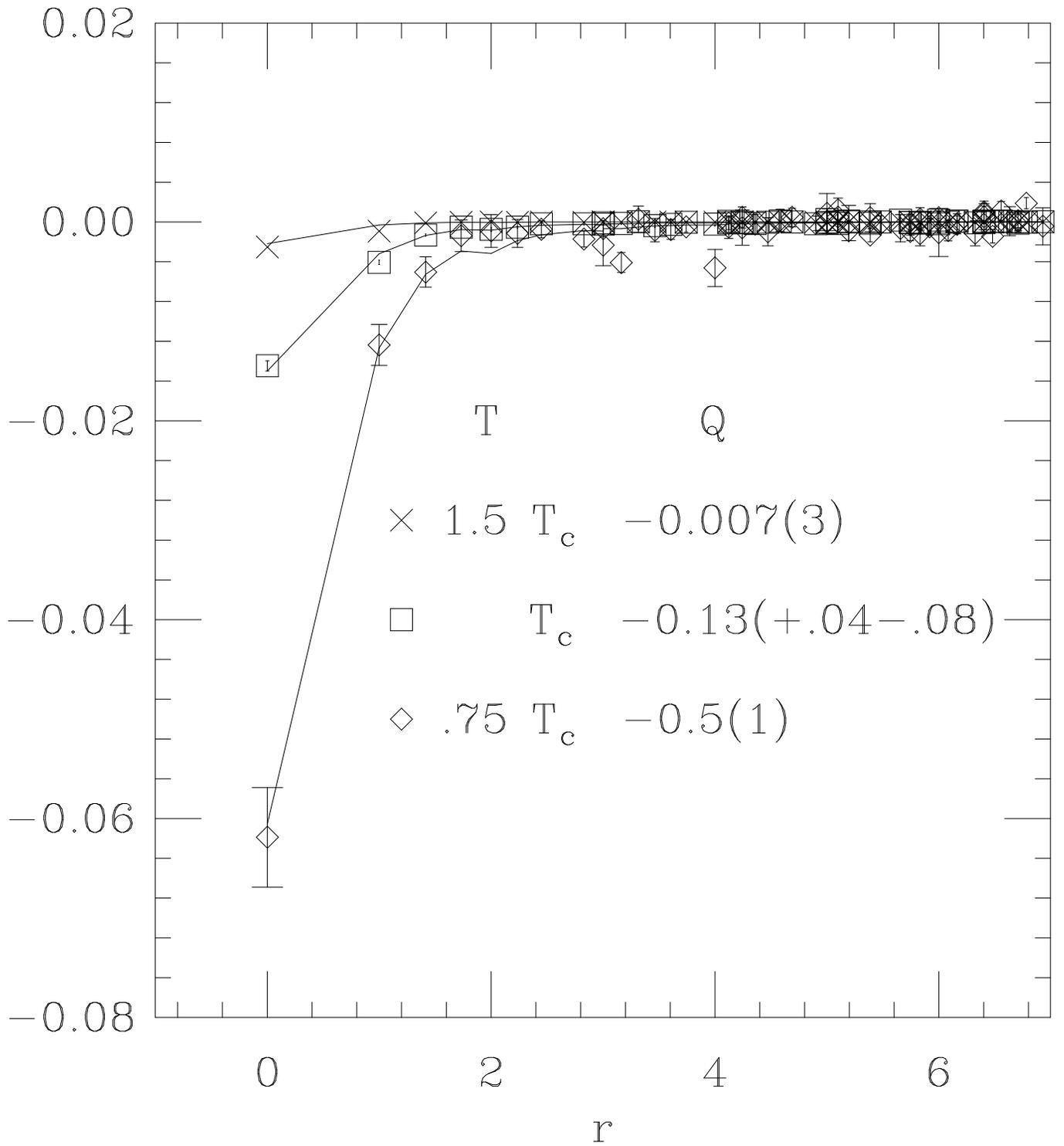
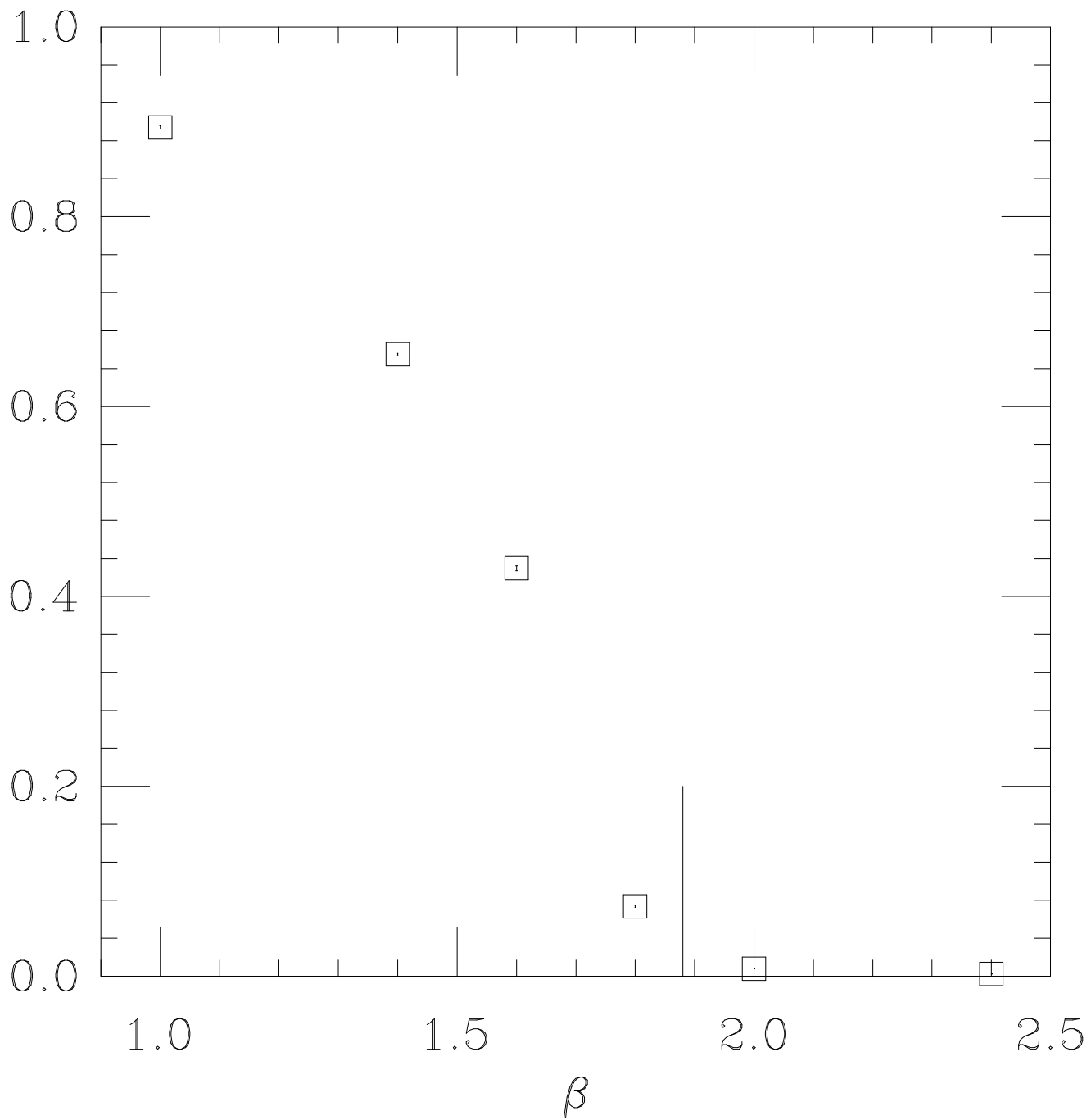
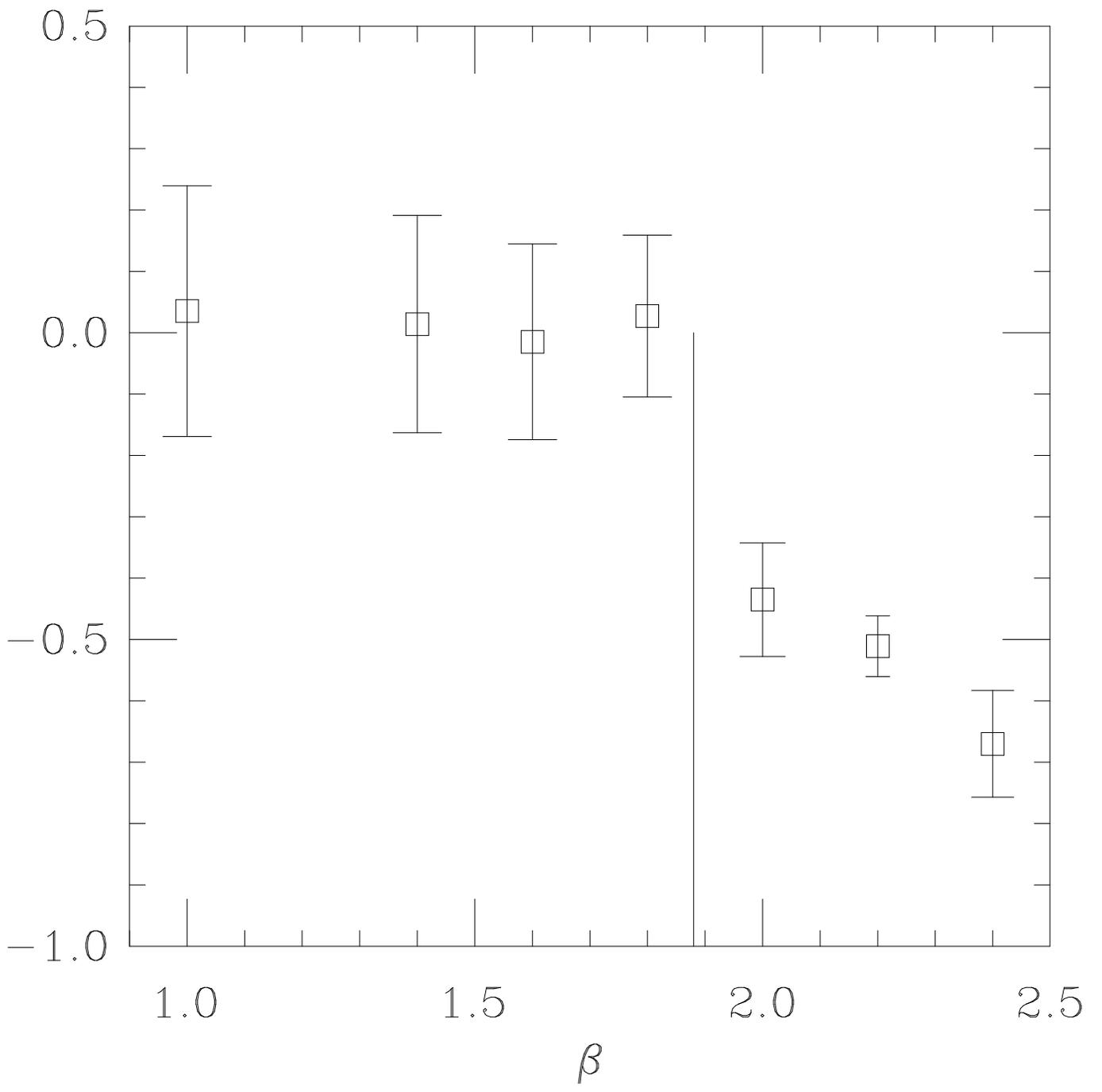
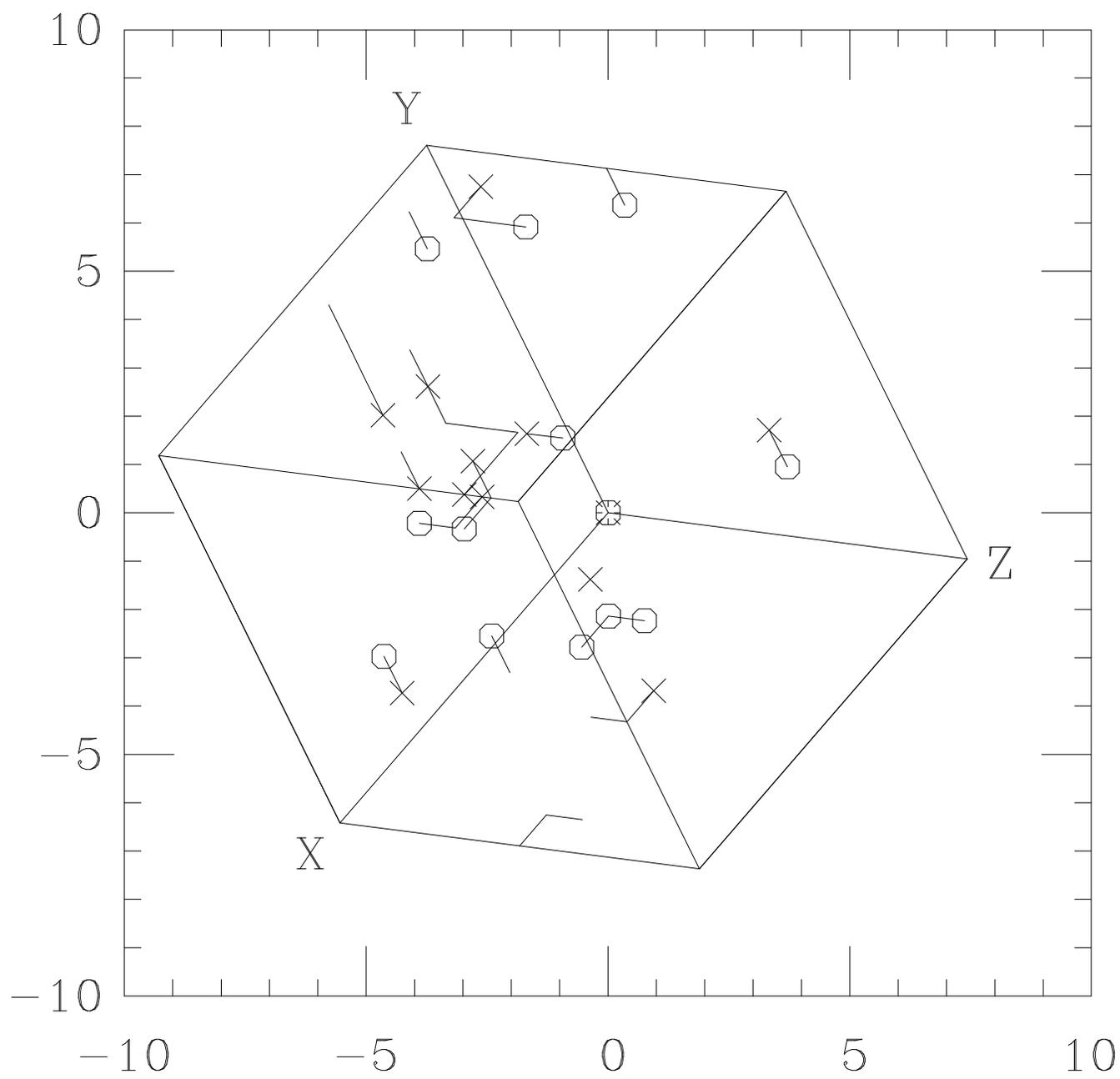
\end{document}